\documentclass[twoside,twocolumn,english]{revtex4}
\usepackage[T1]{fontenc}
\usepackage[latin1]{inputenc}

\makeatletter

\newcommand{\noun}[1]{\textsc{#1}}

\usepackage{babel}
\makeatother
\begin{document}

\title{Relativistic hydrodynamics with sources for cosmological K-fluids}

\author{Alberto Díez-Tejedor%
\footnote{wtbditea@lg.ehu.es%
} and Alexander Feinstein%
\footnote{wtpfexxa@lg.ehu.es%
}}

\address{Dpto. de Física Teórica, Universidad del País Vasco, Apdo. 644, 48080,
Bilbao, Spain.}

\begin{abstract}
We consider hydrodynamics with non conserved number of particles and
show that it can be modeled with effective fluid Lagrangians which
explicitly depend on the velocity potentials. For such theories, the
{}``shift symmetry'' $\phi\rightarrow\phi+$const. leading to the
conserved number of fluid particles in conventional hydrodynamics
is globally broken and, as a result, the non conservation of particle
number appears as a source term in the continuity equation. The particle
number non-conservation is balanced by the entropy change, with both
the entropy and the source term expressed in terms of the fluid velocity
potential. Equations of hydrodynamics are derived using a modified
version of Schutz's variational principle method. Examples of fluids
described by such Lagrangians (tachyon condensate, k-essence) in spatially
flat isotropic universe are briefly discussed. 
\end{abstract}
\maketitle

\section{INTRODUCTION}

It is well known that complex physical phenomena can be often modeled
with good accuracy by an effective theory. One such effective macroscopic
model, for example, is the hydrodynamical model of Landau \cite{landau},
which has had a considerable success in explaining certain features
of the collisions of highly relativistic nuclei \cite{bjorken,cooper}.
The universe, the most complex of all the physical systems, is in
general successfully modeled by an isentropic perfect fluid. Hydrodynamic
language, back in high regard, is now invoked to describe non-trivial
field theories \cite{jackiw}.

In cosmology, as mentioned above, the perfect fluid description, despite
the generic complexity of the system, works fine. One of the usual
assumptions in the conventional hydrodynamical description of the
universe is that the universe expands adiabatically. Closely related
to it is the assertion that the so-called mass, or particle number
conservation, holds. Yet, one can imagine a universe where creation
or destruction of particles takes place. This may happen due to the
time variation and inhomogeneities of the gravitational field itself,
not to discard a more speculative possibility of a universe filled
with white and black holes where particles suddenly appear or disappear.
What kind of an effective hydrodynamics would then describe such a
universe?

There are several ways to approach the problem of the universe where
particles are created or annihilated. If this happens due to quantum
processes, then presumably the most direct approach would be to consider
the quantization of the matter fields on a curved background using
the machinery of the quantum field theory \cite{Birrel}, and then
evaluating the back-reaction of the created fields on the classical
geometry. The promising direction within this approach is the study
of stochastic gravity \cite{hu}.

It is possible, though, that for some reason, one is not interested
in the detailed description of the particle creation (destruction)
mechanism. Then one would be trying to model the effects of the microscopic
processes by a kind of an effective macroscopic model. In hadron-hadron
collision theory \cite{bjorken,cooper}, such an effective macroscopic
model is the Landau's hydrodynamical description.

In the framework of cosmology with non conserved number of particles,
a possible macroscopic description was put forward by Prigogine et
al \cite{prigogine}, and later generalised by Calvao et al \cite{calvao}
some years ago. In this approach, the creation of particles is considered
in the context of thermodynamics of open systems. What follows then,
roughly speaking, is that an extra negative {}``viscous'' pressure
term appears in the energy-momentum tensor to account for the created
particles.

Yet, there exists a more {}``economic'' and elegant way to describe
particle creation (annihilation) without a change in the form of the
energy-momentum tensor, and without introducing an extra pressure
term . To introduce a source term into the particle number conservation
equation it is sufficient to allow entropy flow. The change in the
particle number allowed by the continuity equation will then come
at the expense of the entropy change.

In this paper we are interested in exploring a Lagrangian formulation
of relativistic hydrodynamics with non conserved number of particles.
In the conventional variational approach to relativistic hydrodynamics
developed by Schutz \cite{schutz}, the action does not depend on
the velocity potential, but rather is a functional of its derivatives.
This, in turn, maintains the symmetry $\phi\rightarrow\phi+$const.
which allows particle number conservation. Here, we propose a Lagrangian
formulation for the equations of hydrodynamics, where the Lagrangian,
to break globally the symmetry leading to the particle number conservation,
depends not only on the derivatives, but on the velocity potential
itself.

We propose to modify Schutz's original Lagrangian \cite{schutz,schutz2,brown},
by introducing sources and sinks in the continuity equation, modeled
by a velocity potential dependent function. Our formulation is matemathically
selfconsistent, in that it gives the right set of hydrodynamical equations.
Physically, the fluid Lagrangians we consider have connection to matter
described by the rolling tachyon condensate \cite{sen} or by the
K-essence \cite{armendariz,armentesis}.

\section{THE HYDRODYNAMICS WITH PARTICLE NUMBER VARIATION\label{sec:2}}

We start by assuming that we deal with simple thermodynamical systems
(fluids), which are characterised by a fundamental equation of the
form $U=U\left[S,V,N\right]$, where all the variables have their
usual meanings, and we use $k_{B}=c=8\pi G=1$. Assuming the standard
thermodynamic relations, one may show that such a system is completely
specified by the energy density function $\rho(n,s)$ and the system's
size $V$: \[
U\left[S,V,N\right]=V\rho(n,s),\]
 where $n$ is the particle number density and $s$ is the entropy
per particle. We can write the first law of thermodynamics as \begin{equation}
d\rho=hdn+nTds,\label{eq:priley}\end{equation}
 where $h$ is the enthalpy per particle. Assuming further that the
particle number in the system is not conserved, the equations of hydrodynamics
take the following form: \begin{equation}
T_{\quad;\mu}^{\mu\nu}=0,\label{eq:consener}\end{equation}
\begin{equation}
\left(nu^{\mu}\right)_{;\mu}=\psi,\label{eq:varpati}\end{equation}
 where $T^{\mu\nu}$ stands for the usual stress-energy tensor of
a perfect fluid: \begin{equation}
T^{\mu\nu}=\left(\rho+p\right)u^{\mu}u^{\nu}+pg^{\mu\nu}.\label{eq:energia}\end{equation}
 Here $p$ and $\rho$ are the pressure and the energy density of
the fluid respectively, and $u^{\mu}$ is the four-velocity field
($u_{\mu}u^{\mu}=-1$). The equation (\ref{eq:consener}) represents
the conservation of the energy-momentum tensor, whereas (\ref{eq:varpati})
is the continuity equation with the source ($\psi>0$) or the sink
($\psi<0$) term for the particles. We must further specify the equation
of state $\rho=\rho(n,s)$, along with the source term $\psi$, which
we take to have the form $\psi=\psi(n,s)$. To close this system of
equations we add the first law of thermodynamics (\ref{eq:priley}).
The equations (\ref{eq:consener}), (\ref{eq:varpati}) and (\ref{eq:priley})
form a self consistent field theory describing a fluid with particle
number variation in terms of five macroscopic (or Eulerian) variables
($n,s,u^{\mu}$).

To obtain a more intuitive form of these equations it is convenient
to project the energy conservation equation (\ref{eq:consener}) along
and, in the direction perpendicular, to the four-velocity. The parallel
projection ($u_{\mu}T_{\quad;\nu}^{\mu\nu}=0$), after the balance
equation (\ref{eq:varpati}) and the thermodynamical relations have
been substituted, gives the following continuity equation: \begin{equation}
s_{,\mu}u^{\mu}=-\frac{h\psi}{nT}.\label{eq:vars}\end{equation}
 This equation was first given, in a somewhat different form, by Schutz
and Sorkin \cite{schutz2}. One can appreciate how the change in the
number of particles ($\psi$) is accompanied by a change in the entropy
per particle ($u^{\mu}s_{,\mu}\neq0$). The fluid flow no longer follows
lines of constant $s$, as it happens in the conventional hydrodynamics
when no source is present ($\psi=0$). The projection perpendicular
to the four-velocity gives ($P_{\mu\alpha}T_{\quad;\nu}^{\mu\nu}=0$,
with $P_{\mu}^{\nu}\equiv u^{\nu}u_{\mu}+\delta_{\mu}^{\nu}$): \[
\left(\rho+p\right)u_{\alpha;\nu}u^{\nu}=-p_{,\nu}P_{\alpha}^{\nu},\]
 which is the relativistic Euler equation. The last two equations
are completely equivalent to the eqs. (\ref{eq:consener}) and (\ref{eq:varpati}).

The variation rate of the number of particles $N$ and the total entropy
$S$ of the fluid may still be written in a more suggestive way: \begin{equation}
\frac{dN}{d\tau}=V\psi,\quad\frac{dS}{d\tau}=-\frac{\mu}{T}\frac{dN}{d\tau},\label{eq:NS}\end{equation}
 where $\mu=h-sT$ is the chemical potential. From the first of these
two equations we see that the sign of the source term determines as
to whether the particles are created or annihilated. The other equation
describes the variation of the entropy, whose change is determined
by both, the sign of the chemical potential and the source term.

\section{THE ACTION PRINCIPLE\label{sec:AN-ACTION}}

The relativistic perfect fluid action functionals were developed by
Taub \cite{taub1} and Schutz \cite{schutz}. Here we follow closely
Schutz's velocity potential formalism \cite{schutz}. In the case
of the conventional hydrodynamics, where no particle creation takes
place ($\psi=0$), one starts with the following action \cite{schutz,schutz2,brown}:
\[
S=\int d^{4}x\left\{ -\sqrt{-g}\rho(n,s)+J^{\mu}\left(\phi_{,\mu}+s\theta_{,\mu}+\beta_{A}\alpha_{\:,\mu}^{A}\right)\right\} ,\]
 with $A$ taking the values $1,2$ and $3$. Here, $\phi$ and $\theta$
are Lagrange multipliers introduced to satisfy the particle number
and the entropy conservation constraints respectively. One further
assumes the existence of Lagrangian coordinates $\alpha^{A}$ which
label the flow lines, and consequently introduces the $\beta_{A}$
potentials in form of Lagrange multipliers. $J^{\mu}$ is the particle
number current-density, defined as $J^{\mu}\equiv\sqrt{-g}nu^{\mu}$.
The expression for the current permits to write the particle number
density as $n=\left|J\right|/\sqrt{-g}$. To include the gravity as
a dynamical field into the picture, one adds, as usual, the Einstein-Hilbert
term to the above action. The variables in which the action is formulated
are, therefore: $g^{\mu\nu}$, $J^{\mu}$, $\phi$, $s$, $\theta$,
$\beta_{A}$ and $\alpha^{A}$. Starting with this action, one derives
both the hydrodynamical equations of motion and the energy-momentum
tensor for the fluid \cite{brown}.

We now consider the action principle for the hydrodynamics described
in the previous section. For this purpose, we put forward the following
action \cite{futuro}:\[
S=\int d^{4}x\left\{ -\sqrt{-g}\rho(n,s)+J^{\mu}\left(\phi_{,\mu}+\beta_{A}\alpha_{\:,\mu}^{A}\right)\right\} ,\]
 where now the entropy per particle $s$ is not an independent variable
any more. We assume $s=s(\phi)$, so that the {}``shift symmetry''
$\phi\rightarrow\phi+$const. present in the Schutz's original action
is globally broken. We have also suppressed the term $J^{\mu}s\theta_{,\mu}$
in the action, since now we do not expect the entropy per particle
$s$ to conserve. To justify physically the functional dependence of
the entropy on the velocity potential (apart from the fact that such
an action leads to the equations of motion we expect) we suggest that
since the non-conservation of the particles via the equation (\ref{eq:vars})
leads to the entropy change, on one hand, and that we model the particle
non-conservation by the symmetry breaking with the $\phi$-dependent
term in the action on the other, it looks reasonable to introduce
this dependence in the entropy term. One must bear in mind, however,
that due to the particular parametrisation $s(\phi)$, the particle
variation rate is neither arbitrary, nor generic, yet we find it sufficiently
general for the purposes of the physics we are interested in. With
this in mind, one may show \cite{futuro} that the equations of hydrodynamics
as well as the form of the energy momentum tensor of the section \ref{sec:2}
can be recovered from the above action.

Introducing the equations of motion back into the action we obtain
the on-shell expression \[
S_{on-shell}=\int d^{4}x\sqrt{-g}p,\]
 which coincides with the on-shell expression of Schutz for conventional
hydrodynamics. Thus, we have that the Lagrangian for the hydrodynamics
with particle non-conservation may still be given by the pressure
of the fluid.

\section{THE IRROTATIONAL FLOW \label{sec:IRROTATIONAL}}

We now assume the fluid flow to be irrotational. We also note, that
although until now we have expressed the action in terms of $\rho(n,s)$,
it is often convenient to use, with the help of the usual thermodynamic
relations, a different parametrisation of the action \cite{brown}.
In the context of the irrotational flow it will be more convenient
to work with the equation of state $p=p(h,s)$.

First, let us see what happens in the conventional case when the particle
number is conserved. In this case we have an isentropic fluid ($s=$const.)
and the action may be expressed as \[
S=\int d^{4}x\sqrt{-g}\left\{ p\left(\left|V\right|\right)-\left(\frac{\partial p}{\partial h}\right)_{s}\left[\left|V\right|-\frac{V^{\mu}\varphi_{,\mu}}{\left|V\right|}\right]\right\} ,\]
 where we have defined the current $V^{\mu}\equiv hu^{\mu}$, and
the subindex $s$ refers to the fact that the partial derivative $\left(\partial p/\partial h\right)$
is evaluated at constant $s$. The variables are $g^{\mu\nu}$, $V^{\mu}$
and $\varphi$, and the following equations of motion result: \begin{equation}
u_{\mu}=-h^{-1}\varphi_{,\mu},\label{eq:din1}\end{equation}
\begin{equation}
\left(nu^{\mu}\right)_{;\mu}=0,\label{eq:din2}\end{equation}
 with the energy-momentum tensor given by \begin{equation}
T^{\mu\nu}=\frac{2}{\sqrt{-g}}\frac{\delta S}{\delta g_{\mu\nu}}=\left(\frac{\partial p}{\partial h}\right)_{s}hu^{\mu}u^{\nu}+pg^{\mu\nu}.\label{eq:momento}\end{equation}
 Comparing the last equation with the equation (\ref{eq:energia})
allows to define the pressure and the energy density of the fluid
as: $p=p$ and $\rho=nh-p$ (with $n=\left(\partial p/\partial h\right)_{s}$).
The pressure and the density defined via the stress-energy tensor
coincide with their usual thermodynamical definitions. The equation
(\ref{eq:din1}) is the expression of the fact that the fluid flow
is irrotational, whereas the equation (\ref{eq:din2}) is the particle
number conservation equation.

The identity $u_{\mu}u^{\mu}=-1$ and the equation (\ref{eq:din1}),
lead to the following expression for the enthalpy: \begin{equation}
h=\sqrt{-\varphi_{,\mu}\varphi^{,\mu}}.\label{eq:entalpia}\end{equation}
 To make contact with the now popular K-essence cosmology \cite{armendariz,armentesis},
we write the action on-shell as \begin{equation}
S_{on-shell}=\int d^{4}x\sqrt{-g}F(X),\label{eq:action}\end{equation}
 where we define: \begin{equation}
X\equiv-\frac{1}{2}\varphi_{,\mu}\varphi^{,\mu}=\frac{h^{2}}{2},\label{eq:def1}\end{equation}
\begin{equation}
p(h)=p\left(\sqrt{2X}\right)\equiv F(X),\label{eq:def2}\end{equation}

Therefore, if one has an irrotational fluid where the number of particles
is conserved, it is described by the hydrodynamics derived from the
Lagrangian (\ref{eq:action}), which depends only on the derivatives
of the velocity potential defined by the equation (\ref{eq:din1}).
Moreover, the conservation equation (\ref{eq:din2}), is just the
Euler-Lagrange equation derived from the action (\ref{eq:action}):
\[
-\left(nu^{\mu}\right)_{;\mu}=\left[F'(X)\varphi^{,\mu}\right]_{;\mu}=0,\]
 where we have used $n=\left(\partial p/\partial h\right)_{s}=hF'(X)$,
and the prime stands for the derivative of the function with respect
to its argument. For completeness, we give the expression for the
density and the pressure of the fluid in terms of the variable $X$:
\begin{equation}
p=F(X),\quad\rho=2XF'(X)-F(X).\label{eq:presion}\end{equation}
 These expressions are known in the context of K-field as purely kinetic
K-field \cite{armendariz,armentesis,Scherrer}.

Typically, in cosmology, one uses an equation of state $p=f(\rho)$
to describe an isentropic fluid. To obtain an action in the form (\ref{eq:action})
describing such a fluid, one only has to express the energy density
as $\rho=f^{-1}(p)$, insert the latter expression into the equation
of energy density (\ref{eq:presion}), and obtain the differential
equation for $F,$ $f^{-1}(F)=2XF'-F$. This gives then for $F(X)$:

\begin{equation}
\int^{F}\frac{dF^{*}}{f^{-1}(F^{*})+F^{*}}=\ln\left[CX^{1/2}\right],\label{eq:ecF}\end{equation}
 where $C$ is an arbitrary integration constant. The equation (\ref{eq:ecF})
establishes how to pass from the standard hydrodynamical description
of an isentropic irrotational perfect fluid ($p=f(\rho))$, to the
language of an action principle (\ref{eq:action}). Put differently,
the purely kinetic K-field, is interpretable in terms of an isentropic
perfect fluid with an equation of state which can be easily put into
the form $p=p(\rho)$. Thus, \textit{any solution to the Einstein's
field equations with the energy momentum tensor of the irrotational
perfect fluid with the equation of state $p=p(\rho)$ is by default
interpretable as a solution for the purely kinetic K-fluid}.

We now consider the irrotational flow where the number of particles
is not conserved. In this case, the action can be expressed as \cite{futuro}:\[
S=\int d^{4}x\sqrt{-g}\left\{ p\left(\left|V\right|,s\right)-\left(\frac{\partial p}{\partial h}\right)_{n}\left[\left|V\right|-\frac{V^{\mu}\phi_{,\mu}}{\left|V\right|}\right]\right\} \]
 along with $s=s(\phi)$. The variables still are $g^{\mu\nu}$, $V^{\mu}$
and $\phi$, and the equations of motion that follow are\begin{equation}
u_{\mu}=-h^{-1}\phi_{,\mu},\label{eq:din3}\end{equation}
\begin{equation}
\left(nu^{\mu}\right)_{;\mu}=-nT\frac{ds}{d\phi}.\label{eq:din4}\end{equation}
 In the last equation we have used $-nT=\left(\partial p/\partial s\right)_{h}$.
The form of the energy-momentum tensor is left unchanged and is given
by the equation (\ref{eq:momento}). The equation (\ref{eq:din3})
expresses again the fact that the flow is irrotational, and the continuity
equation (\ref{eq:din4}), if we define the particle creation rate
as \begin{equation}
\psi\equiv-nT\frac{ds}{d\phi},\label{eq:crea}\end{equation}
 is the balance equation (\ref{eq:varpati}). We hence model the creation
of particles through the function $s=s(\phi)$. Using the property
$u_{\mu}u^{\mu}=-1$ and the equation (\ref{eq:din3}), we obtain
the equation (\ref{eq:entalpia}), and now the on-shell action becomes:
\begin{equation}
S_{on-shell}=\int d^{4}x\sqrt{-g}L(\phi,X),\label{eq:action2}\end{equation}
 where we have used the equation (\ref{eq:def1}) and have defined
\begin{equation}
p(h,s)=p\left(\sqrt{2X},s(\phi)\right)\equiv L(\phi,X).\label{eq:ladensity}\end{equation}
 We thus have succeeded in giving the action for the irrotational
fluid flow where number of particles is not conserved in terms of
the scalar velocity potential and its derivatives. Moreover, the continuity
equation of the fluid (\ref{eq:din4}), using $n=\left(\partial p/\partial h\right)_{s}=h\partial L/\partial X$,
$u^{\mu}=-h^{-1}\phi^{,\mu}$ and $\psi=-nTds/d\phi=\partial L/\partial\phi$
becomes the Euler-Lagrange equation for the action (\ref{eq:action2}):
\begin{equation}
\left[\frac{\partial L}{\partial X}\phi^{,\mu}\right]_{;\mu}+\frac{\partial L}{\partial\phi}=0.\label{eq:ecdin}\end{equation}
 We finally express the pressure and the density of the fluid in terms
of the scalar field: \begin{equation}
p=L(\phi,X),\quad\rho=2X\frac{\partial L(\phi,X)}{\partial X}-L(\phi,X).\label{eq:denpres}\end{equation}

\section{K-FLUID}

A special case arises when the fluid has a separable equation of state
$p(h,s)=f(s)g(h)$. In this case, the action takes the form\begin{equation}
S=\int d^{4}x\sqrt{-g}K(\phi)F(X),\label{eq:landensity1}\end{equation}
 with definitions $F(X)\equiv g(\sqrt{2X})$ and $K(\phi)\equiv f\left(s(\phi)\right)$.
The entropy per particle $s$ can be then expressed as a function
of the potential term $K(\phi)$:\begin{equation}
s=f^{-1}\left[K(\phi)\right],\label{eq:s}\end{equation}
 and the equation (\ref{eq:denpres}) permits to express the pressure
and energy as: \begin{equation}
p=K(\phi)F(X),\quad\rho=K(\phi)\left[2XF'(X)-F(X)\right].\label{eq:pres1}\end{equation}
 The above expressions are analogous to factorisable K-field theories
\cite{armendariz,armentesis}, and we therefore refer to these fluids
as K-fluids. The case in which there is no particle creation (purely
kinetic K-fluid) is obtained with $K(\phi)=$const.

One usually assumes without lost of generality that $K(\phi)>0$ ($f(s)>0$),
while $F(X)$ may be either positive or negative, allowing for tensions
instead of pressure. Yet, we want to have a positive energy density,
we therefore must have\begin{equation}
2XF'(X)-F(X)\geq0.\label{eq:con1}\end{equation}
 In addition, the particle number density $n$ must also be positive,
so we need \begin{equation}
F'(X)\geq0\label{eq:con2}\end{equation}
 and \begin{equation}
\textnormal{sgn}\left[f'(s)\right]=-\textnormal{sgn}\left[p\right]\label{eq:con3}\end{equation}
 to have positive temperature. One may further define the sound speed
in a usual way: \[
c_{s}^{2}=\left(\frac{\partial p}{\partial\rho}\right)_{s}=\frac{F'(X)}{2XF''(X)+F'(X)}\]
 (cf \cite{Mukhanov}). For ordinary fluids one usually also imposes
$0\leq c_{s}^{2}\leq1$, and therefore using (\ref{eq:con2}), we
have: \begin{equation}
F''(X)\geq0.\label{eq:con4}\end{equation}

We therefore refer to K-hydrodynamics, or K-fluid for short, as to
an irrotational fluid with an equation of state $p(h,s)=f(s)g(h)$
and a particle variation rate given by $\psi(h,s)=k(s)g(h)$. The
particle variation rate is further parametrised by $s=s(\phi)$, where
$\phi$ is the velocity potential of this irrotational fluid, and
the peculiar functional form of the particle production rate $\psi(h,s)$
is a consequence of this parametrisation choice. The action for the
K-fluid is given by the equation (\ref{eq:landensity1}). We impose
positivity of the energy density (\ref{eq:con1}), the particle number
density (\ref{eq:con2}) and the temperature (\ref{eq:con3}), but
are less stringent with the pressure, though one may always impose
the positivity of the pressure as well. The fluid flow is stable as
long as (\ref{eq:con4}) holds.

We can now express the fluid parameters in terms of the scalar field.
The particle number density and particle production rate take the
form:\begin{equation}
n=K(\phi)\sqrt{2X}F'(X),\label{eq:particle1}\end{equation}
\begin{equation}
\psi=F(X)K'(\phi),\label{eq:rate}\end{equation}
 and consequently the sign of the derivative of $K$ defines as to
whether the creation or annihilation of particles takes place. When
$K(\phi)=$const., the expression $N=Vn$ is the Noether's charge
associated with the {}``shift symetry'' $\phi\rightarrow\phi+$const.
of the action. These expressions above can be written in terms of
the action without the explicit knowledge of the function $f(s)$.
However, to evaluate the entropy per particle (\ref{eq:s}), total
entropy $S=Vns$ and temperature \begin{equation}
T=\frac{-f'(s)}{n}F(X),\label{eq:tem}\end{equation}
 one must know the form of $f(s)$. Some examples will be given in
the following section.

With the above hydrodynamical interpretations, let us look for a moment
at the K-fluids where the number of particles is conserved. The entropy
per particle is then a constant, say $s_{0}$, and therefore the equation
of state has the form $p=p(h)$. This is an isentropic fluid characterised
by the function $F(X)$ and the constant $f(s_{0})$. The action for
the fluid becomes: \begin{equation}
S=\int d^{4}x\sqrt{-g}f(s_{0})F(X),\label{eq:ac}\end{equation}
 where the function $F(X)$ is given by (\ref{eq:ecF}) subject to
the conditions (\ref{eq:con1}), (\ref{eq:con2}) and (\ref{eq:con4}),
while $f(s)$ must verify (\ref{eq:con3}). The Lagrangian (\ref{eq:ac}),
up to a non-essential multiplicative constant, is the Lagrangian for
the purely kinetic K-field \cite{Scherrer}, for which we have defined
the pressure, the energy density (\ref{eq:pres1}), the entropy per
particle (\ref{eq:s}) $s=s_{0}$, the particle number density (\ref{eq:particle1})
and the temperature (\ref{eq:tem}).

To close this section we give the dynamical equation (\ref{eq:ecdin})
in the case of the factorisable K-fluid theory: \begin{equation}
\nabla_{\mu}\left[K(\phi)F'(X)\phi^{,\mu}\right]+K'(\phi)F(X)=0.\label{eq:dyngen}\end{equation}

\section{PARTICULAR EXAMPLES}

Let us consider some particular examples. We start by specifying the
following equation of state: \begin{equation}
p(h,s)=e^{\mp s}g(h),\label{eq:ecstate}\end{equation}
 where we have $-$ for $p>0$ and $+$ for $p<0$, in accordance
with (\ref{eq:con3}). We see that this equation of state is of the
form described in the previous section. One must further specify the
function $s(\phi)$ in terms of the particle creation rate, so that
the entropy per particle (\ref{eq:s}) can be expressed as a function
of the potential: \begin{equation}
s=\mp\ln\left[K(\phi)\right].\label{eq:s1}\end{equation}
 For the entropy per particle to be positive, one should impose $0<K(\phi)<1$
($K(\phi)>1$) for $p>0$ ($p<0$). With the equation of state (\ref{eq:ecstate}),
the temperature of the fluid becomes\[
T=\frac{-1}{n}\frac{\partial p}{\partial s}=\frac{\left|p\right|}{n}.\]
 Note, that this expression for the temperature coincides with the
expression one would have for a typical fluid composed of non-interacting
physical particles (generalized to negative pressures), and is a consequence
of the choice we made for the equation of state (\ref{eq:ecstate}).
In terms of the field we have the particle number density (\ref{eq:particle1}),
particle rate production (\ref{eq:rate}), and with this choice of
$f(s)$ we can compute the entropy per particle (\ref{eq:s1}) and
the temperature (\ref{eq:tem}) \begin{equation}
T=\frac{\left|F(X)\right|}{\sqrt{2X}F'(X)}.\label{eq:temperature1}\end{equation}
 If we consider the case where the particle number remains constant,
the action for the fluid becomes \begin{equation}
S=\int d^{4}x\sqrt{-g}e^{\mp s_{0}}F(X),\label{eq:acsin}\end{equation}
 where the function $F(X)$ is evaluated from (\ref{eq:ecF}).

\textbf{Example 1}: Fluid with constant adiabatic index $p=w\rho$
($w=\textnormal{const.}$). We have $\rho=f^{-1}(F)=F/w$. >From (\ref{eq:ecF})
we obtain: \begin{equation}
F(X)=\pm X^{\frac{1+w}{2w}},\label{eq:F(X)}\end{equation}
 where the sign $+$ corresponds to $w>0$, while the sign $-$ corresponds
to the case $-1\leq w<0$, after the constraints (\ref{eq:con1})
and (\ref{eq:con2}) have been applied %
\footnote{The case of pressureless dust ($w=0$) should be treated separately.
In this case it is convenient to work with the equation of state $\rho=\rho(n)$.%
}. In the case of the stable flow, the constraint (\ref{eq:con4})
imposes the positivity of the pressure together with $0<w\leq1$.
Such a fluid is then described by the action \[
S=\int d^{4}x\sqrt{-g}e^{-s_{0}}X^{\frac{1+w}{2w}},\]
 with $0<w\leq1$, and where \[
p=e^{-s_{0}}X^{\frac{1+w}{2w}},\quad\rho=\frac{e^{-s_{0}}}{w}X^{\frac{1+w}{2w}},\]
 \[
n=e^{-s_{0}}\frac{(1+w)}{\sqrt{2}w}X^{\frac{1}{2w}},\quad T=\frac{\sqrt{2}w}{1+w}X^{\frac{1}{2}}.\]
 In a spatially flat FRW universe ($ds^{2}=-dt^{2}+a^{2}(t)\left[dr^{2}+r^{2}\left(d\theta^{2}+\sin^{2}\theta d\varphi^{2}\right)\right]$),
solving the dynamical (\ref{eq:dyngen}) and Friedmann ($3H^{2}=\rho$)
equations , we easily recover $X(a)\propto a^{-6w}$ and $a(t)\propto t^{2/3(1+w)}$.

Now, let us turn to the theories where the number of particles is
not conserved. The kind of action we have is (\ref{eq:landensity1}),
sticking to the factorisable theories. A typical potential to use
would be the well studied $K(\phi)\propto1/\phi^{2}$ \cite{armendariz,armentesis,Feinstein},
due to the fact that it leaves one with solutions with constant enthalpy
per particle in spatially flat isotropic universes. Yet, if we would
use the equation of state (\ref{eq:ecstate}) we would certainly run
into trouble because of the restrictions on the function $K(\phi)$.
There are two ways to circumvent this problem: either consider a different
equation of state, or a different potential. If we stick to the above
equation of state, then for example the following potentials $K(\phi)=A\cosh\phi$
with $A\geq1$ and $K(\phi)=A\exp(-\phi^{2})+B$, with $A,B>0$ and
$A+B<1$, will do. Now, the only advantage of using the equation of
state (\ref{eq:ecstate}) is that the temperature is given as the
ratio of the pressure to the particle number density. The next simplest
choice for an equation of state would be the one for which the entropy
function $f(s)$ is a power-law. We thus take \begin{equation}
p(h,s)=s^{b}g(h),\label{eq:eqstate2}\end{equation}
 where $b$ is an arbitrary constant such that sgn$(b)=-$sgn$(p)$
to satisfy (\ref{eq:con3}). The entropy then, according to (\ref{eq:s}),
is\begin{equation}
s=\left[K(\phi)\right]^{\frac{1}{b}},\label{eq:s2}\end{equation}
 and is compatible with the potentials of the form $K(\phi)\propto1/\phi^{2}$.
The particle number density and the particle rate production in terms
of the field are given by (\ref{eq:particle1}) and (\ref{eq:rate}),
whereas the temperature now takes the form \begin{equation}
T=\frac{\left|bF(X)\right|}{\sqrt{2X}F'(X)}\left[K(\phi)\right]^{-\frac{1}{b}}.\label{eq:temperature2}\end{equation}

We will now look at two {}``similar'' fluids in a spatially flat
FRW universe, but with different properties with respect to the particle
number conservation. In the first case, the fluid is isentropic with
a conserved particle number and provokes interest in both field theory
\cite{jackiw} and cosmology \cite{Chap}. The second case represents
the same fluid where the number of particles is not conserved and
has a Lagrangian of the form of Sen's tachyon condensate \cite{sen},
which has recently become of considerable interest in cosmology \cite{Feinstein,tachyon}.

\textbf{Example 2}: Fluid with equation of state $p=-A/\rho$, $A=$const.$>0$
(Chaplygin gas)\emph{.} In this case we have $\rho=f^{-1}(F)=-1/F$.
Inserting this in (\ref{eq:ecF}), we obtain, up to some unessential
constants: \[
F(X)=\pm\sqrt{1\pm X}.\]
 We assume the constraints (\ref{eq:con1}) and (\ref{eq:con2}) hold,
and we are left therefore with negative pressure: \begin{equation}
F(X)=-\sqrt{1-X}\label{eq:F(X)1}\end{equation}
 with $0\leq X\leq1$, and there is no problem with the constraint
(\ref{eq:con4}), indicating that the flow is stable. We can think
of such a fluid as a fluid with the equation of state \begin{equation}
p(h,s)=-s^{b}\sqrt{1-\frac{h^{2}}{2}}\label{eq:eqstate3}\end{equation}
 in which the number of particles is conserved ($b=\textnormal{const.>0}$).
The action is then \[
S=-\int d^{4}x\sqrt{-g}\left(s_{0}\right)^{b}\sqrt{1-X},\]
 and therefore \[
p=-\left(s_{0}\right)^{b}\sqrt{1-X},\quad\rho=\frac{\left(s_{0}\right)^{b}}{\sqrt{1-X}},\]
 \[
n=\left(s_{0}\right)^{b}\sqrt{\frac{X}{2\left(1-X\right)}},\quad T=\frac{b}{s_{0}}\sqrt{\frac{2}{X}}\left(1-X\right).\]
 Solving the field equation (\ref{eq:dyngen}) in a spatially flat
FRW model, one obtains \[
X(a)=\frac{1}{1+Ba^{6}},\]
 where $B$ is an integration constant. From here one may evaluate
all the hydrodynamical parameters in terms of the scale factor, arriving
to the unusual result that the temperature of the Chaplygin gas rises
with the expansion. This basically happens due to the negative pressure
of the fluid \cite{Lima}. One can further solve, as well, the Friedmann
equation to find the behaviour of the scale factor as a function of
time \cite{Chap}.

\textbf{Example 3}: Tachyon condensate. The possibility of fluid description
of tachyon condensate in bosonic and supersymmetric string theories
discovered by Sen \cite{sen} has motivated a considerable amount
of work studying the consequences of the rolling tachyon in cosmology
\cite{Feinstein,tachyon}. Here we are interested to look at tachyon
condensate action in the light of the formalism developed above as
a fluid where the number of particles is not conserved. The action
for the K-fluid with the form of the tachyon condensate is \cite{Feinstein,sen}\[
S=-\int d^{4}x\sqrt{-g}K(\phi)\sqrt{1-X}.\]
 We can think of the above action as one describing a fluid with equation
of state (\ref{eq:eqstate3}) in which the particle rate production
is modeled by (\ref{eq:s2}). From this action we can reed off: \[
p=-K(\phi)\sqrt{1-X},\quad\rho=\frac{K(\phi)}{\sqrt{1-X}},\]
 \[
n=K(\phi)\sqrt{\frac{X}{2(1-X)}},\quad T=\frac{b}{\left[K(\phi)\right]^{\frac{1}{b}}}\sqrt{\frac{2}{X}}\left(1-X\right).\]
 It is simple to obtain particular cosmological solutions for such
a fluid if one assumes a spatially flat isotropic cosmology and a
potential of the form $K(\phi)=\beta/\phi^{2}$ with $\beta>0$ \cite{Feinstein}.
Solving Einstein's equations one finds for the velocity potential
$\phi(t)$ and the scale factor of the universe $a(t)$: \begin{equation}
\phi(t)=\sqrt{\frac{4}{3n}}t,\quad a(t)=t^{n},\label{eq:scale}\end{equation}
 where $n$ is a constant given in terms of the parameter of the potential
$n=n(\beta)$. Therefore the parameter $\beta$, the slope of the
potential, defines the particle creation rate as well as different
expansion rate. For these particular solutions the enthalpy per particle
of the fluid (\ref{eq:entalpia}) remains constant. This is in contrast
with the case of Chaplygin gas, where the entropy per particle was
constant.

We can use the expressions (\ref{eq:NS}) to evaluate the increase
of the number of particles and entropy of the system in a time interval
$\Delta t$. For the toy models with equation of state of the form
(\ref{eq:eqstate2}), and a particle creation rate modeled by $K(\phi)=\beta/\phi^{2}$,
we obtain the following expressions in a spatially flat FRW universe:
\[
\Delta N(t_{1},t_{2})=-\frac{8\pi\beta}{3}F(X)\int_{t_{1}}^{t_{2}}\left(\frac{a}{\phi}\right)^{3}dt,\]
 \[
\Delta S(t_{1},t_{2})=-\frac{8\pi\beta^{\frac{1+b}{b}}}{3}F(X)\left[1+\frac{2XF'(X)}{bF(X)}\right]\int_{t_{2}}^{t_{1}}\frac{a^{3}}{\phi^{\frac{2+3b}{b}}}dt.\]
 For the tachyon-like model, taking into consideration (\ref{eq:scale}),
we have \[
\Delta N\propto t^{3n-2},\quad\Delta S\propto t^{\frac{3nb-2\left(b+1\right)}{b}}.\]
 Since we must impose $X<1$ for the action to be well-defined, one
has $n>2/3$, and, interestingly enough, this implies that the particles
are created in such a universe. The creation rate is best 
visualised  by  the expression

\[
\frac{1}{N}\frac{dN}{dt}=\frac{1}{\sqrt{2X}}\frac{\frac{d}{d\phi}\left[\ln K(\phi)\right]}{\frac{d}{dX}\left[\ln F(X)\right]}.\]
For the above tachyon example we  readily find that the creation rate fades with time
as $t^{-1}$.

We see that the fluid we have is the same as in Chaplygin gas (has
the same equation of state), but the production of particles changes
the evolution of the universe. Changing the particle creation rate
one changes the expansion rate of the model.

\section{CONCLUSIONS}

In this paper we have considered a Lagrangian approach to a Relativistic
Hydrodynamics in which the number of particles is not conserved. The
particle number non-conservation is modeled by introducing an explicit
velocity potential dependent term into the fluid Lagrangian. In doing
so, the usual shift symmetry of the action is broken, resulting in
the appearance of a source term in the continuity equation. The conservation
equation derived from the stress-energy tensor indicates that the
particle number non-conservation must be balanced by an entropy flow.
Both the entropy flow and the change in the particle number are expressed
as function of the velocity potential. Although such a description
is valid for a general flow, we concentrate on the purely potential
fluid motion without vorticity, to make contact with some modern theories
used for the description of matter in the universe.

By identifying the K-essence field variable $2X$ with the \emph{square}
of the enthalpy per particle $h$ we identify the K-field theory and
the hydrodynamical Lagrangians we look at. In the case of purely kinetic
K-essence, we observe that this theory is identical to the isentropic
perfect fluid, and give a `dictionary' (\ref{eq:ecF}) as to how to
pass from the usual description in cosmology in terms of the equation
os state $p=p(\rho)$ to the K-theory Lagrangians of the form $F(X)$.
On a formal level, therefore, the purely kinetic K-essence is no `big
news', but rather a simple conventional hydrodynamics in a disguise.

The non-conventional hydrodynamics (K-hydrodynamics), the one analogous
to the K-essence with the potential term, is rather more involved.
First, one must interpret such a hydrodynamics as a flow where the
number of particles is not conserved. This, in turn, leads to a change
in the entropy per particle, as well as to a global entropy flow.
The fluid now is not isentropic and to give an hydrodynamical description
the two equations of state $p=f(s)g(h)$ and $\psi=\psi(s,h)$ must
be specified. We have found \cite{futuro} that our parametrisation
works for source terms of the form $\psi(s,h)=k(s)g(h)$, i.e. the
source term must be separable in functions of entropy and enthalpy,
and the enthalpy function must be the same as the one which appears
in the pressure. This restricts the generality of the approach, nevertheless,
it is of direct application to the K-essence-like cosmologies.

We have finally cosidered several examples of fluids with both conserved
and non-conserved number of particles in the context of spatially
flat isotropic universe. The telling example is the comparison of
Chaplygin gas on one hand and a K-fluid with the form of tachyon condensate
on the other. In the first case one deals with an isentropic perfect
fluid where the number of particles is conserved. The peculiarity
of this example is that the temperature of the gas rises up with the
expansion. The second example represents a fluid with the same equation
of state, but with the number of particles (entropy per particle)
not conserved. It is interesting, however, that for special creation
rates, those with the potential $K=\beta/\phi^{2}$ with $\beta>0$,
the enthalpy per particle rather than the entropy remains constant
in the course of the expansion. We also observe that in such a universe
creation rather than destruction of particles takes place.

From a technical point of view it appears that the velocity potential/$X$-variable
formalism is quite useful to study the dynamics of the cosmological
models. It would be interesting in the future to obtain and study
some of the K-fluid type Lagrangians obtained as effective theories
derived from fundamental interactions. The work in this direction
is in progress and will be presented elsewhere.

\begin{acknowledgments}
We are grateful to Manuel Valle for enlighting discussions. A.D.T.
is grateful to Ruth Lazkoz and Sanjay Jhingan for encouragement. A.D.T.
work is supported by the Basque Government predoctoral fellowship
BFI03.134. This work is supported by the Spanish Science Ministry
Grant 1/MCYT 00172.310-15787/2004. 
\end{acknowledgments}


\begin{thebibliography}{10}
\bibitem{landau}L.D. Landau, Izv. Akad. Nauk SSSR \textbf{17}, 51 (1953); L.D. Landau
and S.Z. Belenkij, Usp. Phys. Nauk \textbf{56},309 (1955). 
\bibitem{bjorken}J.D. Bjorken, Phys. Rev. D \textbf{27}, 140 (1983). 
\bibitem{cooper}F. Cooper, G. Frye and E. Schonberg, Phys. Rev. D \textbf{11}, 192
(1975), F. Cooper and D. Sharp, Phys. Rev. D \textbf{12}, 1123 (1975) 
\bibitem{jackiw}R. Jackiw, arXiv: physics/0010041. R. Jackiw, V.P. Nair, S.-Y. Pi,
A.P. Polychronakos, arXiv: hep-ph/0407101. 
\bibitem{Birrel}N.D. Birrel and P.C.W. Davies, Quantum Fields in Curved Space (Cambridge
Monographs on Mathematical Physics). 
\bibitem{hu}B.L. Hu and E. Verdaguer, Class. Quant. Grav. \textbf{20}, R1 (2003). 
\bibitem{prigogine}I. Prigogine, J. Geheniau, E. Gunzig and P. Nardone, Gen. Rel. Grav.
\textbf{21}, 767 (1989), Proc. Natl. Acad. Sci. USA \textbf{85}, 7428
(1988). 
\bibitem{calvao}M. Calvao, J. Lima and I. Waga, Phys. Lett. A \textbf{162}, 223 (1992). 
\bibitem{schutz}B. Schutz, Phys. Rev. D \textbf{2}, 2762 (1970). 
\bibitem{schutz2}B. Schutz and R. Sorkin, Annals Phys. \textbf{107}, 1 (1977). 
\bibitem{brown}J.D. Brown, Class. Quant. Grav. \textbf{10}, 1579 (1993). 
\bibitem{taub1}A. Taub, Phys. Rev. \textbf{94}, 1468 (1954); A. Taub, Commun. Math.
Phys. \textbf{15}, 235 (1969). 
\bibitem{sen}A. Sen, JHEP \textbf{0204}, 048 (2002). 
\bibitem{armendariz}C. Armendariz-Picon, T. Damour and V. Mukhanov, Phys. Lett. B \textbf{458},
209 (1999); C. Armendariz-Picon, V. Mukhanov and P. J. Steinhardt,
Phys. Rev. D \textbf{63}, 103510 (2001), Phys. Rev. Lett. \textbf{85},
4438 (2000). 
\bibitem{armentesis}C. Armendariz-Picon, Ph.D. Thesis, http://edoc.ub-unimuenchen.de/archive/00000186/
. 
\bibitem{futuro}A. Díez-Tejedor and A. Feinstein (in preparation). 
\bibitem{Scherrer}Scherrer, Phys. Rev. Lett. \textbf{93} 011301 (2004); Frolov, Phys.
Rev. D \textbf{70}, 061501 (2004).
\bibitem{Mukhanov}J. Garriga and V.F. Mukhanov, Phys. Lett. B \textbf{458}, 219 (1999). 
\bibitem{Feinstein}A. Feinstein, Phys. Rev. D \textbf{\noun{66}}, 063511 (2002); T.
Padmanabhan, Phys. Rev. D \textbf{66} (2002) 021301. 
\bibitem{Chap}A.Yu. Kamenshchik, U. Moschella and V. Pasquier, Phys. Lett B \textbf{511},
265 (2001). 
\bibitem{Lima}J.A.S. Lima and J.S. Alcaniz, Phys. Lett. B \textbf{600}, 191 (2004). 
\bibitem{tachyon}G.W. Gibbons, Phys. Lett. B \textbf{537}, 1 (2002); M. Fairbairn and
M.H.G. Tytgat, Phys. Lett. B \textbf{546}, 1 (2002); S. Mukohyama,
Phys. Rev. D \textbf{66}, 024009 (2002); D. Choudhury, D. Ghoshal,
D. P. Jatkar and S. Panda, Phys. Lett. B \textbf{544}, 231 (2002);
G. Shiu and I. Wasserman, Phys. Lett. B \textbf{541}, 6 (2002); L.
Kofman and A. Linde, JHEP \textbf{0207}, 004 (2002); M. Sami, Mod.
Phys. Lett. A \textbf{18}, 691 (2003); A. Mazumdar, S. Panda and A.
Perez-Lorenzana, Nucl. Phys. B \textbf{614}, 101 (2001); J.C. Hwang
and H. Noh, Phys. Rev. D \textbf{66}, 084009 (2002); Y.S. Piao, R.G.
Cai, X.M. Zhang and Y. Z. Zhang, Phys. Rev. D \textbf{66}, 121301
(2002); J.M. Cline, H. Firouzjahi and P. Martineau, JHEP \textbf{0211},
041 (2002); G.N. Felder, L. Kofman and A. Starobinsky, JHEP \textbf{0209},
026 (2002); S. Mukohyama, Phys. Rev. D \textbf{66}, 123512 (2002);
M.C. Bento, O. Bertolami and A.A. Sen, Phys. Rev. D \textbf{67}, 063511
(2003); J.G. Hao and X.Z. Li, Phys. Rev. D \textbf{66}, 087301 (2002);
C.J. Kim, H.B. Kim and Y.B. Kim, Phys. Lett. B \textbf{552}, 111 (2003);
T. Matsuda, Phys. Rev. D \textbf{67}, 083519 (2003); A. Das and A.
DeBenedictis, Gen. Rel. Grav. \textbf{36}, 1741 (2004); Z.K. Guo,
Y.S. Piao, R.G. Cai and Y.Z. Zhang, Phys. Rev. D \textbf{68}, 043508
(2003); M. Majumdar and A.C. Davis, Phys. Rev. D \textbf{69}, 103504
(2004); S. Nojiri and S.D. Odintsov, Phys. Lett. B \textbf{571}, 1
(2003); E. Elizalde, J.E. Lidsey, S. Nojiri and S.D. Odintsov, Phys.
Lett. B \textbf{574}, 1 (2003); D.A. Steer and F. Vernizzi, Phys.
Rev. D \textbf{70}, 043527 (2004); V. Gorini, A.Y. Kamenshchik, U.
Moschella and V. Pasquier Phys. Rev. D \textbf{69}, 123512 (2004);
L.P. Chimento, Phys. Rev. D \textbf{69}, 123517 (2004); M.B. Causse,
arXiv:astro-ph/0312206; B.C. Paul and M. Sami, Phys. Rev. D \textbf{70},
027301 (2004); G.N. Felder and L. Kofman, Phys. Rev. D \textbf{70},
046004 (2004); J.M. Aguirregabiria and R. Lazkoz, Mod. Phys. Lett.
A \textbf{19}, 927 (2004); E.J. Copeland, M.R. Garousi, M. Sami and
S. Tsujikawa, arXiv:hep-th/0411192.\end{thebibliography}
\end{document}